\documentclass[a4paper]{article}

\usepackage{INTERSPEECH2022}
\usepackage{graphicx,xcolor}
\usepackage{slashbox}
\usepackage{multirow}
\usepackage{arydshln}
\usepackage[noend]{algorithmic}
\usepackage{algorithm} 

\newcommand{\ra}[1]{\renewcommand{\arraystretch}{#1}}

\title{Streaming Multi-Talker ASR with Token-Level Serialized Output Training}
 
 \name{Naoyuki Kanda$^1$, Jian Wu$^1$, Yu Wu$^2$, Xiong Xiao$^1$, Zhong Meng$^1$, Xiaofei Wang$^1$, Yashesh Gaur$^1$,\\Zhuo Chen$^1$, Jinyu Li$^1$, Takuya Yoshioka$^1$}
 \address{
   $^1$Microsoft Cloud+AI, USA \hspace{3mm}$^2$Microsoft Research Asia, China} 
\email{\{nakanda,wujian,yuwu1,xioxiao,zhme,xiaofewa,yagaur,zhuc,jinyli,tayoshio\}@microsoft.com}

\begin{document}

\maketitle
\begin{abstract}
This paper proposes a token-level serialized output training (t-SOT), a novel framework for streaming multi-talker automatic speech recognition (ASR). Unlike existing streaming multi-talker ASR models using multiple output branches, the t-SOT model has only a single output branch that generates recognition tokens (e.g., words, subwords) of multiple speakers in chronological order based on their emission times. A special token that indicates the change of ``virtual'' output channels is introduced to keep track of the overlapping utterances. Compared to the prior streaming multi-talker ASR models, the t-SOT model has the advantages of less inference cost and a simpler model architecture. Moreover, in our experiments with LibriSpeechMix and LibriCSS datasets, the t-SOT-based transformer transducer model achieves the state-of-the-art word error rates by a significant margin to the prior results. For non-overlapping speech, the t-SOT model is on par with a single-talker ASR model in terms of both accuracy and computational cost, opening the door for deploying one model for both single- and multi-talker scenarios.
\end{abstract}

\section{Introduction}

Speech overlaps are ubiquitous in human-to-human conversations.
For example, it was reported that 
6--15\% of
speaking time was overlapped in meetings 
\cite{ccetin2006analysis,yoshioka2018recognizing}.
The overlap rate can be
even higher for daily conversations \cite{kanda2019guided,barker2018fifth,watanabe2020chime}.
Nevertheless, most of the current automatic speech recognition (ASR) 
systems are designed to
transcribe non-overlapping audio,
and even a short overlap period 
significantly hurts
the ASR accuracy \cite{chen2020continuous,raj2020integration}.
A traditional approach for recognizing the overlapping speech
is to apply speech separation, 
followed by applying ASR for each separated  signal.
While shown to be  effective (e.g., \cite{kanda2019guided,yoshioka2019advances,chen2021continuous}),
such an approach makes the entire system complicated
and hard to optimize for the best ASR accuracy.

To further improve the accuracy for the multi-talker scenario,
various approaches were proposed 
to
 train an ASR model
that can directly 
recognize multiple utterances from the multi-talker audio.
A popular approach is 
to use a neural network that has multiple output branches
to generate transcriptions for overlapping speakers
(e.g., \cite{yu2017recognizing,seki2018purely,chang2019end,chang2019mimo,tripathi2020end}), where the model is often trained with permutation invariant training~\cite{hershey2016deep,isik2016single,yu2017permutation}.
Another recently proposed approach  is 
serialized output training (SOT) \cite{kanda2020sot}, which
uses a model that has only a single output branch.
In SOT, the single output branch generates transcriptions for multiple speakers
one after another, where the speaker-wise transcriptions are interleaved by  a special separator token that indicates the speaker change.
The SOT-based ASR model achieved the state-of-the-art (SOTA) word error rate (WER) \cite{kanda2021end,kanda2021large,kanda2021comparative}
for various multi-talker test sets 
including LibriSpeechMix \cite{kanda2020sot}, LibriCSS \cite{chen2020continuous} 
and AMI \cite{carletta2005ami}. 
However, the SOT model assumes the attention-based encoder-decoder (AED) \cite{chorowski2014end,chorowski2015attention}
as a backbone ASR system, which renders
the model usable only for the offline (i.e. non-streaming) inference.

\begin{figure}[t]
  \centering
  \includegraphics[width=1.0\linewidth]{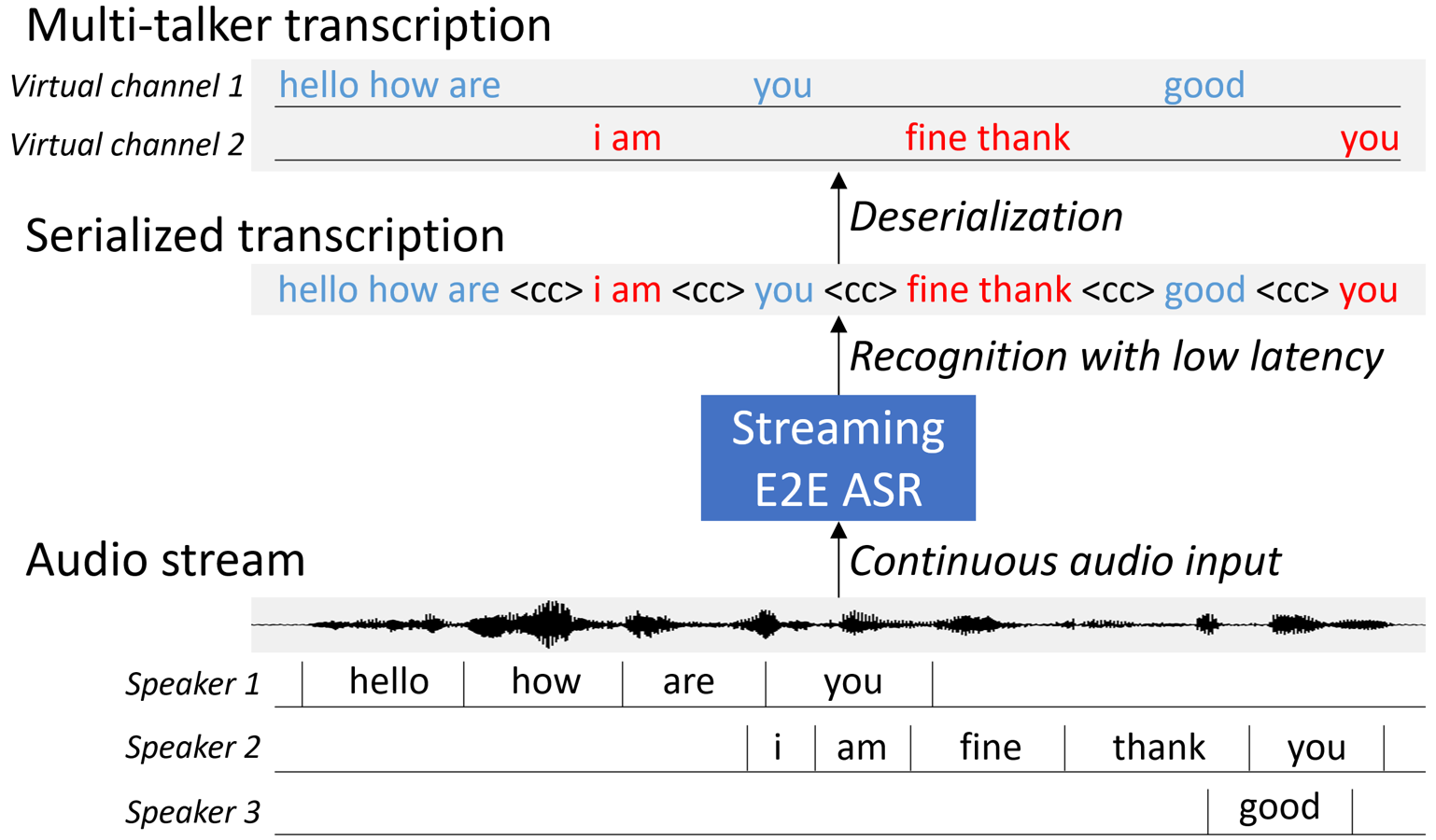}
  \vspace{-7mm}
      \caption{An overview of the token-level serialized output training for a case with up to two concurrent utterances.}
  \label{fig:word_sot}
   \vspace{-5mm}
\end{figure}

\begin{table*}[t]
\ra{1.0}
  \caption{Comparison of representative multi-talker ASR frameworks. A preferable property in each row is presented in {\bf bold font}.}
  \label{tab:framework_comparison}
 \vspace{-3mm}
  \centering
{ \footnotesize
\begin{tabular}{@{}llll@{}}
    \toprule 
 & SOT \cite{kanda2020sot} & SURT \cite{lu2021streaming}, MS-RNN-T \cite{sklyar2021streaming} & t-SOT (proposed method) \\ \midrule
Streaming inference & Not-available & {\bf Available} & {\bf Available} \\ 
Inference cost on decoder & {\bf Same with single-talker model} & $K$-times$^\ast$ of single-talker model & {\bf Same with single-talker model}   \\
Max concurrent utterances & {\bf Unlimited} & Pre-defined & Pre-defined \\
Speaker counting & {\bf Available} & Not-available & Not-available \\ \midrule
Model architecture & {\bf Same with single-talker model} & Multiple output branches & {\bf Same with single-talker model} \\
Restriction on ASR-type & Restricted to AED-based ASR& {\bf No restriction} & {\bf No restriction} \\ \midrule
Accuracy on non-overlapping audio$^\diamond$  & {\bf Good}  & Bad & {\bf Good} \\ 
Accuracy on overlapping audio$^\diamond$      & {\bf Good}  & Fair& {\bf Good} \\ \bottomrule
  \end{tabular}
  \\\hspace{-0mm}$^\ast$ $K$ is the number of output branches of the model.
  \;\;$^\diamond$ Good/Fair/Bad are judged based on the results in Tables \ref{tab:librispeechmix_summary}--\ref{tab:libricss_summary}.
  }
  \vspace{-5mm}
\end{table*}

A few recent studies explored the 
streaming multi-talker ASR problem to
transcribe each spoken word with a low latency even for overlapping speech.
Streaming unmixing and recognition transducer (SURT) \cite{lu2021streaming} and 
multi-speaker recurrent neural network transducer (MS-RNN-T) \cite{sklyar2021streaming} were
concurrently proposed based on a similar idea, where the model
has two output branches to generate two simultaneous transcriptions for overlapping speech.
However,  their reported WERs still lagged far behind
 the SOTA result of the offline SOT-model.

In this paper, we present 
token-level serialized output training (t-SOT),
a novel streaming multi-talker ASR framework.
With t-SOT, words spoken by overlapping 
speakers are generated by a single output branch in a chronological order based on their emission times.\footnote{The idea of emission-time-based serialization was concurrently proposed in \cite{chang22arxiv}.
However, their speaker tracking method is different from ours, and their experiments were limited to non-streaming models.}
A special
token indicating ``virtual'' output channels is used to
keep track of transcriptions for overlapping speakers.\footnote{While several prior works proposed to insert special tokens in a transcription  \cite{watanabe2017language,inaguma2018end,el2019joint, kanda2020sot}, our work is the first to propose a special token for transcribing {\it overlapping} utterances in a {\it streaming} fashion.}  
Compared to the prior streaming
multi-talker ASR models \cite{lu2021streaming,sklyar2021streaming}, 
the t-SOT model has advantages of
less inference cost and a simpler model architecture. 
Moreover,
our experimental results with 
 LibriSpeechMix and LibriCSS datasets show that the transformer transducer (TT) \cite{zhang2020transformer} trained with 
t-SOT framework  achieves new SOTA results by 
a significant margin to the prior results including the offline models.

\section{Streaming multi-talker ASR with t-SOT}

\subsection{Basic t-SOT for up to two concurrent utterances}
We first explain the basic idea of t-SOT 
by assuming 
that
  the input audio contains overlapping speech of up to two concurrent utterances 
 as shown in Fig. \ref{fig:word_sot}. 
 Note that the audio may contain utterances of any number of speakers 
 as long as the number of \textit{concurrent} utterances 
is less than or equal to two.

Our idea is to serialize transcriptions for multiple speakers by sorting the recognition tokens of all speakers in a chronological order. 
The algorithm to generate the serialized transcription for a training sample 
is described in Algorithm \ref{algo:wSOT_2}.
Suppose we have a training audio sample 
with a time- and speaker- annotated transcription. 
By joining all words in all speakers' transcriptions, 
we can represent such transcription as
a sequence $\mathcal{T}=(w_i, e_i, s_i)_{i=1}^{N}$,
where
$w_i$ is the $i$-th token (e.g., word, subword) in the joined sequence,
 $e_i$ is the emission time (i.e. end time) of $w_i$, and $s_i$ is
 the 
 speaker of $w_i$ (line 1).
We first sort $\mathcal{T}$ based on $e_i$ in an ascending order (line 2).
The serialized transcription $\mathcal{W}$ is initialized as a list that contains the first token $w_1$ in $\mathcal{T}$ (line 3).
We then iteratively append $w_i$ for $i=2,...,N$ while
inserting a special ``channel change'' token $\langle \mathrm{cc}\rangle$ 
if the speaker of the consecutive two tokens in $\mathcal{T}$ are different (lines 4--7).
The resultant $\mathcal{W}$ is the serialized transcription of up to two concurrent utterances (line 8),
which is exemplified in the ``serialized transcription'' in Fig. \ref{fig:word_sot}.

We train a streaming end-to-end (E2E) ASR model based on the audio samples and the corresponding serialized transcriptions.
Any streaming ASR models can be used, including connectionist temporal classification
\cite{graves2006connectionist}, RNN-T \cite{graves2012sequence}, 
and TT \cite{zhang2020transformer},
without changing the structure from the conventional single-talker  model. 
During inference,
a sequence of tokens including $\langle cc \rangle$ is
generated in a streaming fashion, which can then be reformatted to
two virtual output channels by switching the channel index at the recognized
$\langle \mathrm{cc} \rangle$ position as shown in the ``Deserialization'' process in Fig. \ref{fig:word_sot}.
Note that the first recognized token is always assigned to the first virtual channel.

\begin{figure}[t]
\vspace{-3.5mm}
\begin{algorithm}[H]
 \caption{Generating a serialized transcription for a training sample with up to two concurrent utterances}
\footnotesize
\label{algo:wSOT_2}
\begin{algorithmic}[1]
\STATE{\textbf{Given} a speech sample with a time- and speaker-annotated transcription $\mathcal{T}=(w_i, e_i, s_i)_{i=1}^N$ where $w_i$ is $i$-th token, and $e_i$ and $s_i$ are the emission time and the speaker of $w_i$, respectively.}
\STATE{Sort $\mathcal{T}$ in ascending order of $e_i$.} 
\STATE{Initialize a list of tokens $\mathcal{W}\leftarrow[w_1]$.} 
\FOR{$i \;\mathrm{in}\; 2,...,N$}
        \IF{$s_i \neq s_{i-1}$}
        \STATE{Append $\langle \mathrm{cc}\rangle$ to $\mathcal{W}$.}
        \ENDIF
        \STATE{Append $w_i$ to $\mathcal{W}$.}
\ENDFOR
\STATE{\textbf{return} $\mathcal{W}$}
\end{algorithmic}
\end{algorithm}
\vspace{-10mm}
\end{figure}

\subsection{Generalized t-SOT for up to $M$ concurrent utterances}

The idea of t-SOT can be easily extended to deal with up to $M$ concurrent utterances by 
introducing $M$
special tokens $\{\langle \mathrm{cc}_1 \rangle,...,\langle \mathrm{cc}_M \rangle\}$,
where $\langle \mathrm{cc}_m \rangle$ indicates 
that the subsequent tokens will be placed to the $m$th virtual channel.
A complete algorithm to generate the serialized transcription is shown in Appendix A. 
Note that we conducted experiments with only up to two concurrent utterances 
because overlaps of more than two speakers 
are rare in many practical scenarios  \cite{ccetin2006analysis}, and 
it seems difficult
for humans to read more than two simultaneous transcriptions in real time.

\subsection{Comparison to prior works}
\label{sec:advantage}
Table \ref{tab:framework_comparison} summarizes the differences between t-SOT and representative  multi-talker ASR methods.
Compared with SOT \cite{kanda2020sot} which can be used only for non-streaming inference,
a t-SOT model can recognize each spoken word in a streaming fashion.
The t-SOT model is also advantageous because the model architecture is not restricted to 
the AED.
On the other hand, t-SOT has the limitation on the maximum number of concurrent utterances
that the model can recognize.
The t-SOT model also cannot count the number of speakers in the audio segment.

For streaming multi-talker ASR, 
the t-SOT framework has various advantages over
SURT \cite{lu2021streaming} and MS-RNN-T \cite{sklyar2021streaming}.
Firstly, t-SOT requires only a single decoding process 
as with the conventional single-talker ASR
while SURT and MS-RNN-T require to execute the decoder multiple times (i.e., one decoder run for each output branches).
Therefore, the t-SOT model fundamentally requires less computation than
the prior models.
Secondly, the t-SOT model architecture is much simpler than SURT and MS-RNN-T because
the same architecture as the single-talker ASR model can be used without any modification.
 Thirdly, our experimental results showed that the t-SOT-based ASR model achieved
 significantly better WER than the prior SOTA results 
 for both non-overlapping and overlapping speech as detailed in the next section.

\begin{table}[t]
\ra{1.0}
  \caption{WER (\%) for LirbiSpeechMix test set with various non-streaming and streaming multi-talker ASR models.
  The algorithmic latency is shown in the ``Latency'' column, 
  where 
  \underline{160 msec} is our target configuration.
  No language model was used for all results including the ones in prior works.
  }
  \label{tab:librispeechmix_summary}
  \vspace{-3mm}
  \centering
 {  \footnotesize
\begin{tabular}{@{}lrrrr@{}}
    \toprule
Model & \multirow{2}{*}{\shortstack[l]{\# of\\param.}} & \multirow{2}{*}{\shortstack[c]{Latency\\(msec)}}& \multicolumn{2}{c}{Test WER}  \\ \cmidrule{4-5}
      & &   & 1spk & 2spk   \\ \midrule
{\it (Non-streaming Models)}&  \\      
PIT LSTM-AED \cite{kanda2020sot}  & 161M$^\dagger$ & $\infty$  & 6.7 & 11.9   \\ 
SOT LSTM-AED \cite{kanda2020joint}  & 136M\hspace{1.2mm} & $\infty$  & 4.5 & 10.3   \\ 
SOT Confomer-AED \cite{kanda2021end}  & 129M\hspace{1.2mm} & $\infty$  &  3.6 & 4.9    \\ \midrule 
{\it (Streaming Models)}&  \\      
LSTM-SURT \cite{lu2021surit}         & 81M$^\dagger$ & 150  &- & 10.3   \\ 
PIT-MS-RNN-T \cite{sklyar2021streaming} & 81M$^\dagger$ & 30 & 7.6 & 10.2  \\ 
Transformer-SURT \cite{lianglu22}    & 85M$^\dagger$ & 1000 &- & 9.1   \\ 
PIT-MS-RNN-T \cite{sklyar2021multi} & 81M$^\dagger$ & 30 & - & 8.8  \\ \hdashline[1pt/2pt]\hdashline[0pt/1pt]
t-SOT TT-18 & 82M\hspace{1.2mm} &  40 & 5.1  & 8.4    \\ 
t-SOT TT-18 & 82M\hspace{1.2mm} & \underline{160} & 4.9 & 6.9    \\ 
t-SOT TT-18 & 82M\hspace{1.2mm} & 640 & 4.2 & 6.2    \\ 
t-SOT TT-18 & 82M\hspace{1.2mm} & 2560 & 3.9 & 5.2    \\ 
t-SOT TT-36 & 139M\hspace{1.2mm} & \underline{160} & 4.3 & 6.2    \\ 
t-SOT TT-36 & 139M\hspace{1.2mm} & 2560 & 3.3 & 4.4    \\ \bottomrule
  \end{tabular}
  }
  \\{\multirow{2}{*}{\shortstack[l]{\footnotesize $^\dagger$ Models with two output branches, for which
  most parameters are\\\hspace{1mm} \footnotesize loaded and used twice in the inference.}}}
  \vspace{0.5mm}
\end{table}

\section{Experiments}
\label{sec:experiments}
We first conducted an experiment
by using the LibriSpeechMix evaluation set \cite{kanda2020sot},
where we limited the number of speakers in each audio segment to up to two.
We then conducted an evaluation with LibriCSS \cite{chen2020continuous} 
where
each long-form audio 
contains many utterances from 8 speakers. 

\subsection{Experiment with LibriSpeechMix}
\subsubsection{Experimental settings}
\label{sec:librispeechmix_setting}
As the first evaluation dataset, we used LibriSpeechMix \cite{kanda2020sot}, 
which is made by mixing up to three utterances randomly sampled from LibriSpeech \cite{panayotov2015librispeech}.
In this work, we used the single-speaker (i.e. non-overlapping) evaluation set
and the two-speaker-mixed evaluation set.
For the two-speaker-mixed evaluation set,
two utterances are mixed with a randomly-determined delay such that
each evaluation sample contains a partial speaker overlap.
We evaluated the WER in the same way as the prior work \cite{kanda2020sot} did.
That is, the recognition hypotheses (= one or two hypotheses after deserialization)
and the references 
are compared for all possible speaker permutations,
and the speaker permutation that produces the minimum number of errors
is selected to calculate the WER.
A hypothesis (or reference) that does not have a corresponding reference (or hypothesis) is
regarded as all insertion (or deletion) errors.

We simulated the training data by randomly mixing up to two utterances
from ``train\_960'' of the LibriSpeech training data.
In the training data generation,
for each sample,
we first selected the number of speakers $S$ 
from \{1, 2\}
with $p$\% and $(100-p)$\%  probabilities for $S=1$ and $2$, respectively, where $p=50$
 unless otherwise stated.
We then randomly selected $S$ utterances from 
``train\_960''.
If $S$ was equal to two, we further randomly sampled the delay $d$ for the second utterance $u_2$
from $\mathrm{Uniform}(0,\mathrm{len}(u_1))$, where 
$\mathrm{len}(u_1)$ is the duration of the first utterance $u_1$.
Utterance $u_1$ and utterance $u_2$, delayed by $d$, 
were then mixed without changing 
the original volumes.
We used the time-alignment generated by 
the Montreal Forced Aligner \cite{mcauliffe2017montreal} 
to generate the serialized transcription for each training sample.\footnote{If one word consisted of multiple subwords,
the emission time was shared for all subwords while keeping the order of subwords.}
To increase the variability of the training data,
we applied 
the speed perturbation \cite{ko2015audio} with the ratios of $\{0.9, 1.0, 1.1\}$,
the volume perturbation with
the ratio between 0.125 to 2.0, 
and the adaptive SpecAugment \cite{park2020specaugment}.
Following \cite{kanda2021large,sklyar2021multi},
we simulated the training data on the fly 
to generate infinite variations 
of the training samples.

For the streaming ASR model,
we used a TT
with a chunk-wise look-ahead proposed in \cite{chen2021developing}. 
The encoder consists of 
2 convolution layers, each of which halves
the time resolution, followed by
a 18-layer or 36-layer transformer with relative positional encoding.
We refer to the model with the 18-layer transformer as ``TT-18'' and the model with
the 36-layer transformer as ``TT-36''.
Each transformer block consists of
a 512-dim multi-head attention with 8 heads 
and a 2048-dim point-wise feed forward layer with Gaussian error linear unit (GELU) activation function.
Our TT's prediction network consists of 
2 layers of 1024-dim long short-term memory (LSTM).
We used 4,000 word pieces plus blank and $\langle \mathrm{cc}\rangle$ tokens
as the recognition units.
The audio input feature is an 80-dim log mel-filterbank extracted every 10 msec.
As proposed in \cite{chen2021developing}, we controlled the algorithmic latency
of the model
based on the chunk size of the attention mask.
The minimum possible latency is 40 msec, which is determined by the time resolution of 
the input feature sequence and the 2 convolutional layers with 2-times subsampling 
in the encoder.
For all models, 
we performed 225K training iterations with
16 GPUs, each of which consumed mini-batches of 12,000 frames.
We used an AdamW optimizer with
a linear decay learning rate schedule with a peak learning rate 
of 1.5e-3 
after 25K warm up iterations.

\begin{table}[t]
\ra{1.0}
  \caption{WER (\%) on LibriSpeechMix 
 for single- and multi-talker models
  with different training configurations. 
  All models have 160 msec of algorithmic latency.}
  \label{tab:compare_to_single_talker}
 \vspace{-3mm}
  \centering
{    \footnotesize
  \tabcolsep = 1.6mm
\begin{tabular}{@{}llllrrrrr@{}}
    \toprule
Model & \multicolumn{2}{c}{Training data}&&  \multicolumn{2}{c}{Dev WER} &&  \multicolumn{2}{c}{Test WER}  \\ \cmidrule{2-3} \cmidrule{5-6} \cmidrule{8-9}
                            & 1spk & 2spk && 1spk & 2spk && 1spk & 2spk \\ \midrule
single-talker TT-18 & 100\% & -    && 4.3 & 63.2 && 4.5 & 63.7    \\ 
t-SOT TT-18         & 67\%  & 33\% && 4.3 & 7.3  && 4.7 & 7.4    \\
t-SOT TT-18         & 50\%  &50\%  && 4.3 & 6.9  && 4.9 & 6.9    \\ \midrule
single-talker TT-36 & 100\% & -    && 3.9 & 62.9 && 4.4 & 63.3     \\ 
t-SOT TT-36         &  67\% & 33\% && 3.8 & 6.4  && 4.4 & 6.3    \\
t-SOT TT-36         &  50\% & 50\% && 3.9 & 6.0  && 4.3 & 6.2    \\ \bottomrule
  \end{tabular}
  }
  \vspace{-5mm}
\end{table}

\begin{table*}[t]
\ra{1.0}
  \caption{WER (\%) on the monaural LibriCSS test set in the continuous input evaluation setting.
  A macro average of WERs is shown in the ``Avg.'' column.
  0L and 0S are 0\% overlap conditions with long and short inter-utterance silences.
  For each overlapping condition, the best WER with streaming ASR models is shown in {\bf bold font}, and the best number among all ASR models are shown with \underline{underline}. }
  \label{tab:libricss_summary}
 \vspace{-3mm}
  \centering
{ \footnotesize 
\begin{tabular}{@{}llrrrrrrr@{}}
    \toprule
System & Algorithmic latency &\multicolumn{7}{c}{WER (\%) for different overlap ratio} \\ \cmidrule{3-9}
        &  & 0L & 0S & 10 & 20 & 30 & 40 & Avg.   \\ \midrule
\multicolumn{9}{l}{\it(Non-streaming ASR models with speech separation)} \\
BLSTM-CSS + Hybrid ASR \cite{chen2020continuous}  & 1.2 sec$^\ddagger$ + (utterance length)$^\star$   & 16.3 & 17.6   & 20.9 & 26.1 & 32.6 & 36.1 & 24.9    \\ 
Conformer-CSS + Transformer-AED-ASR w/ LM \cite{chen2021continuous} & 1.2 sec$^\ddagger$ + (utterance length)$^\star$ & \underline{6.1} & 6.9   &  9.1 & 12.5 & 16.7 & 19.3 & 11.8   \\ 
Conformer-CSS + Transformer-AED-ASR w/ LM \cite{wu2021investigation} & 1.2 sec$^\ddagger$ + (utterance length)$^\star$ & 6.4 & 7.5    & 8.4  & 9.4 & 12.4 & 13.2 & 9.6    \\ \midrule
\multicolumn{8}{l}{\it(Streaming ASR models)} \\
SURT w/ DP-LSTM \cite{raj2021continuous}       &  350 msec           & 9.8 & 19.1  & 20.6 & 20.4 & 23.9 & 26.8 & 20.1\\
SURT w/ DP-Transformer \cite{raj2021continuous} & 350 msec            & 9.3    & 21.1  & 21.2 & 25.9 & 28.2 & 31.7 & 22.9\\ \hdashline[1pt/2pt]\hdashline[0pt/1pt]
Single-talker TT-18 & 160 msec    & 7.0 & 7.3 & 14.0 & 20.9 & 27.9 & 34.3 & 18.6 \\
Single-talker TT-36 & 160 msec    & {\bf 6.5} & 6.7 & 13.1 & 20.4 & 27.0 & 34.0 & 18.0 \\ 
t-SOT TT-18 (proposed) & 160 msec & 7.5 & 7.5 & 8.5 & 10.5  & 12.6  & 14.0 & 10.1 \\
t-SOT TT-36 (proposed) & 160 msec & 6.7 & \underline{\bf 6.1} & \underline{\bf 7.5} & \underline{\bf 9.3}  & \underline{\bf 11.6} & \underline{\bf 12.9} & \underline{\bf 9.0} \\\bottomrule
  \end{tabular}
  }
  \\{ \footnotesize
  $^\ddagger$ Latency incurred by CSS. 
  \;\;$^\star$ Latency incurred by VAD and ASR. The average of utterance lengths in the LibriCSS test set is 7.5 sec.}
  \vspace{-5mm}
\end{table*}

\subsubsection{Main results}
\label{sec:main_results}
Table \ref{tab:librispeechmix_summary} shows the comparison of 
our t-SOT model and prior multi-talker ASR models 
on the LibriSpeechMix test set.
We evaluated t-SOT TT models 
with various algorithmic latency and model sizes.
Firstly, 
we observed that the t-SOT TT-18 with only 40 msec algorithmic latency 
already outperformed the results of all prior streaming multi-talker ASR models.
Note that 
even though t-SOT TT-18 has almost the same  number of parameters 
with SURT \cite{lu2021streaming,lu2021surit}  or MS-RNN-T \cite{sklyar2021streaming,sklyar2021multi},
t-SOT is more 
time- and space-efficient 
in the inference because
SURT and MS-RNN-T run decoding twice, once for each of the two output branches.
Secondly, we observed 
a significant WER reduction by 
 increasing algorithmic latency and
 the model size.
In our experiment, enlarging the latency for 4 times (e.g., 160 msec to 640 msec) achieved
a similar level of WER reduction to 
doubling the number of layers (i.e., TT-18 to TT-36). 
Notably, our t-SOT TT-36 with 2560 msec latency 
achieved 3.3\% and 4.4\% of WERs for single-speaker 
and two-speaker-mixed test sets, respectively,
which are
even better than 
the prior SOTA results
by the offline SOT Conformer-AED \cite{kanda2021end}.\footnote{Prior offline SOT Conformer-AED \cite{kanda2021end} was trained with
a finite set of simulated data instead of 
on-the-fly data generation \cite{kanda2021large,sklyar2021multi}, which is most likely the reason of better 
WER by our t-SOT TT. }

 \subsubsection{Comparison with single-talker ASR models}
 \label{sec:exp-comparison-with-single-talker}
%
 
Table \ref{tab:compare_to_single_talker} shows a comparison 
 of various single-talker and 
t-SOT models. 
For the t-SOT model, we tested
different mixes of single- and two-speaker training samples
by changing $p$ in the on-the-fly data generation (Section \ref{sec:librispeechmix_setting}).
Firstly, as we expected, the single-talker TT model showed bad WERs for
two speaker overlapping speech, which proves the necessity of multi-talker ASR modeling.
Secondly, 
we observed 
marginal WER improvement
for the single-speaker evaluation set
when the model observed more single-speaker training samples,
which however came with the cost of WER degradation for the two-speaker-mixed evaluation set.
Thirdly, we observed a slight degradation in single-talker WER for the t-SOT TT-18 compared to the single-talker TT-18 most likely because TT-18 does not have 
sufficient capacity to fully learn the representation of
normal tokens along with the additional $\langle \mathrm{cc}\rangle$ token.
On the other hand, 
t-SOT TT-36 achieved even better results than single-talker TT-36
 for the single-talker evaluation data. 
 Our interpretation for this result is as follows:
Because the overlapping speech in the training data works as a kind of data augmentation,
even the accuracy for non-overlapping speech can be improved
as long as the model has sufficient capacity to absorb the training data variations.
Note that a similar improvement for single-talker test 
set was 
reported for SOT-based ASR \cite{kanda2020sot}.

It is noteworthy that
 the t-SOT model achieved almost the same, and sometimes even better, WER compared
 to the single-talker ASR models for the single-talker evaluation set.
 This property is desirable to deploy one model 
  for both single-talker
and multi-talker application scenarios.
On the contrary,
 severe WER degradation
for the non-overlapping speech was reported for the prior streaming multi-talker ASR models \cite{sklyar2021streaming,raj2021continuous},
 where the authors of \cite{raj2021continuous}
attributed the degradation to duplicated or no hypotheses being generated
from multiple output branches.
We believe 
the t-SOT-based model can largely eliminate this issue
by using only a single output branch.

\subsection{Experiment with LibriCSS}
\subsubsection{Experimental settings}
To evaluate the t-SOT model in a more realistic setting where 
the input audio stream 
contains more than two speakers with up to two concurrent utterances at each time frame,
we conducted an evaluation with LibriCSS \cite{chen2020continuous}.
LibriCSS
 is a set of 8-speaker 
recordings made by playing back ``test\_clean''
of LibriSpeech
in a real meeting room.
The original recording was made with a 7-ch microphone array,
and we used the first channel of the recording
in our experiment.
The recordings are 10 hours long in total, and 
they are categorized by the speaker overlap ratio from 0\% to 40\%.
In each category, there are 10 mini-sessions,
each of which is 10 min long.
We used sessions 1 to 9 (i.e. excluding the session 0) for the evaluation
by following the official data split \cite{chen2020continuous}. 

We trained TT models by initializing the parameters
based on the model trained in the LibriSpeechMix experiment.
We simulated additional training samples consisting of  
$S\sim\mathrm{Uniform}(1,5)$ utterances on the fly.
 Randomly generated room impulse responses and noise were further
 added to simulate the reverberant recordings.
We performed 50K training iterations with
16 GPUs, each of which consumed a mini-batch of 12,000 frames.
The AdamW optimizer was used with
a linear decay learning rate schedule  
starting from the learning rate of 1.5e-4.

We evaluated TT models 
by following ``continuous input evaluation'' setting \cite{chen2020continuous},
where the ASR system was applied to
pre-segmented audio (roughly 1--2 min) without using utterance boundary information.
We simply applied our TT models to the audio stream
without any further segmentation like
voice activity detection (VAD).
By following \cite{chen2020continuous}, 
we evaluated speaker-agnostic
WER (SAgWER) \cite{fiscus2006multiple}.\footnote{
It is highly computational demanding to 
calculate SAgWER 
especially for many hypotheses 
without time alignment information \cite{fiscus2006multiple}.
Due to the difficulty, we don't have a comparable result for 
SOT-based ASR models.
For t-SOT models, we were still able to calculate SAgWER 
since the number of concurrent hypotheses was limited to two.}

\subsubsection{Evaluation results}
The evaluation results are shown in Table \ref{tab:libricss_summary}.
We observed the proposed t-SOT TT model achieved 
significantly better WER for all conditions compared 
to the prior streaming ASR models.
Surprisingly, t-SOT TT-36 with 160 msec latency 
even outperformed the strong prior results based on
the Conformer-based continuous speech separation (CSS) \cite{chen2021continuous}
and an offline Transformer-AED-based ASR \cite{wang2019semantic} with language model (LM) fusion  for 5 out of 6 conditions,
resulted in the new SOTA average WER.
This result strongly indicates the advantage of the end-to-end multi-talker modeling over the approach to combine independent modules. 
It is also noteworthy that the t-SOT models came close to 
the same-sized single-talker TT models' performance for the non-overlapping test conditions (0L and 0S).
The t-SOT TT-36
even outperformed the single-talker TT-36 for non-overlapping conditions on average (6.4\% vs. 6.6\%), 
which is consistent with the result for LibriSpeechMix (Section \ref{sec:exp-comparison-with-single-talker}).

\section{Conclusions}
In this paper, we presented
t-SOT, a novel framework for streaming multi-talker ASR.
Unlike prior streaming multi-talker ASR models,
the t-SOT model has only a single output branch 
that generates recognition tokens from overlapping speakers
in chronological order based on
their emission times.
A special token that indicates the change of the virtual output channels is introduced
to keep track of the overlapping speakers.
We evaluated t-SOT with LibriSpeechMix and LibriCSS
and showed that t-SOT model achieved new SOTA results 
even with streaming inference using a simpler model architecture.

\bibliographystyle{IEEEtran}

\bibliography{mybib}

\newpage
\appendix
\section{Generalized t-SOT for up to $M$ concurrent utterances}

Our proposed algorithm to generate a serialized transcription for a training sample that includes
up to $M$ concurrent utterances is described in Algorithm \ref{algo:wSOT_M}.
Here, we assume the transcription $\mathcal{T}^+$ additionally contains
a Boolean value $b_i\in\{\mathrm{True},\mathrm{False}\}$  that indicates
whether $w_i$  is the last token of an utterance or not (line 1). 
The utterance boundary can be defined
based on the existence of silence region between two tokens from the same speaker, or can be defined semantically by a human transcriber.
We sort $\mathcal{T}^+$ in an ascending order of $e_i$ (line 2) and
initialize  $\mathcal{F}$ (line 3) as  with Algorithm \ref{algo:wSOT_2}.
 We then introduce a speaker-to-channel dictionary $\mathcal{D}$ 
 and a set of unused channels $\mathcal{C}$ to keep track of the used/unused channels
(lines 4--5).
During the iterative procedure (lines 6-17) for appending tokens one by one (line 13),
an appropriate channel change token $m$ 
is selected and inserted such that the same channel is not used by multiple speakers
at the same time (lines 8--13).
If the $i$-th token is the last token of the corresponding utterance (line 15), 
the channel used by speaker $s_i$ is returned to $\mathcal{C}$ (line 16)
while being deleted from $\mathcal{D}$
(line 17).

At the inference time, 
deserialization is
 conducted 
 with the
$M$ virtual output channel, 
where
the output channel is changed to $m$ when $\langle \mathrm{cc}_m\rangle$ is recognized. 
Note that the first recognized token is always assigned to channel 1.

\begin{algorithm}[h]
\caption{Generating a serialized transcription for a training sample with up to $M$ concurrent utterances}
\footnotesize
\label{algo:wSOT_M}
\begin{algorithmic}[1]
\STATE{\textbf{Given} a speech sample with a time- and speaker-annotated transcription $\mathcal{T^+}=(w_i, e_i, s_i, b_i)_{i=1}^N$ where $b_i\in\{\mathrm{True}, \mathrm{False}\}$ indicates if the $w_i$ is the last token of an utterance or not.}
\STATE{Sort $\mathcal{T^+}$ in ascending order of $e_i$.} 
\STATE{Initialize a list of tokens $\mathcal{W}\leftarrow[w_1]$.} 
\STATE{Initialize a speaker-to-channel dictionary $\mathcal{D}[s_1]\leftarrow\langle \mathrm{cc}_1\rangle$.} 
\STATE{Initialize a set of unused channels $\mathcal{C}\leftarrow\{\langle \mathrm{cc}_2\rangle,...,\langle \mathrm{cc}_M\rangle\}$.} 
\FOR{$i\; \mathrm{in}\; 2,...,N$}
        \IF{$s_i \neq s_{i-1}$}
        \IF{$s_i\in \mathrm{key\_of}(\mathcal{D})$}
        \STATE{$m \leftarrow \mathcal{D}[s_i]$.}
        \ELSE
        \STATE{$m \leftarrow \mathrm{pop}(\mathcal{C})$.}
        \STATE{Add entry $\mathcal{D}[s_i]\leftarrow m$.}
        \ENDIF
        \STATE{Append $m$ to $\mathcal{W}$.}
        \ENDIF
        \STATE{Append $w_i$ to $\mathcal{W}$}
        \IF{$b_i\; \mathrm{is}\; \mathrm{True}$}
        \STATE{$\mathcal{C} \leftarrow \mathcal{C} \cup \{\mathcal{D}[s_i]\}$.}
        \STATE{Delete entry $\mathcal{D}[s_i]$}
        \ENDIF
\ENDFOR
\STATE{\textbf{return} $\mathcal{W}$}
\end{algorithmic}
\end{algorithm}

\section{Using t-SOT for single-talker ASR}

Table \ref{tab:inference} shows the WER for LibriSpeechMix where we
intentionally ignored $\langle \mathrm{cc}\rangle$ token at the time of inference 
with t-SOT models.
We simply set zero for the output probability of 
$\langle \mathrm{cc}\rangle$ token.
This evaluation was designed for the use case where
we want to deploy a t-SOT model 
for both single- and multi-talker scenarios
and we have prior knowledge that
the input audio does not contain overlapping speech, which is possible in some applications. 
As shown in the table, if we ignore 
$\langle \mathrm{cc}\rangle$ token in inference,
we achieved an improved WER for the single-talker (i.e. non-overlapping) evaluation set at the  cost of severe degradation 
for the two-speaker-mixed evaluation set.

\begin{table}[t]
\ra{1.0}
  \caption{
  WER (\%) on LibriSpeechMix 
 for single- and multi-talker models with different inference configurations.
All models have 160 msec of algorithmic latency.  
}
  \label{tab:inference}
 \vspace{-3mm}
  \centering
{   \footnotesize
 \tabcolsep = 1.6mm
\begin{tabular}{@{}llrrrrr@{}}
    \toprule
Model & \multirow{2}{*}{\shortstack[l]{Ignore $\langle cc\rangle$\\in inference}}&  \multicolumn{2}{c}{Dev WER}&&  \multicolumn{2}{c}{Test WER}  \\ \cmidrule{3-4}\cmidrule{6-7}
          & & 1spk & 2spk && 1spk & 2spk \\ \midrule
single-talker TT-18 & - & 4.3 & 63.2 && 4.5 & 63.7    \\ 
t-SOT TT-18 & yes & 4.2 & 58.2 &&  4.7  & 58.5     \\ 
t-SOT TT-18 & no  & 4.3 & 6.9 && 4.9 & 6.9    \\ \midrule
single-talker TT-36 & - & 3.9 & 62.9 && 4.4 & 63.3     \\ 
t-SOT TT-36 & yes & 3.9 & 57.9 && 4.2 &  58.4   \\ 
t-SOT TT-36 & no & 3.9 & 6.0 && 4.3  & 6.2    \\ \bottomrule
  \end{tabular}
  
  }
\end{table}

\section{Effect of beam size in inference}

Table \ref{tab:beam_size} shows the effect of beam size in the inference of t-SOT models
based on LibriSpeechMix evaluation set.
We observed that the large beam size such as 16 is especially important to achieve the best WER for overlapping speech.
At the same time, we also observed the beam size of 4 or 8 already achieved a good WER, 
and the beam size is not necessary to be unrealistically large.
Note that we used the beam size of 16 for all other experiments.

\begin{table}[h]
\ra{1.0}
  \caption{Effect of the beam size in inference of t-SOT models.
  WERs (\%) on LibriSpeechMix are shown.
All models have 160 msec of algorithmic latency.
}
  \label{tab:beam_size}
 \vspace{-3mm}
  \centering
{   \footnotesize
\begin{tabular}{@{}lrrrrrr@{}}
    \toprule
Model & \multirow{2}{*}{\shortstack[c]{Beam\\size}}&  \multicolumn{2}{c}{Dev WER}&&  \multicolumn{2}{c}{Test WER}  \\ \cmidrule{3-4}\cmidrule{6-7}
          & & 1spk & 2spk && 1spk & 2spk \\ \midrule
t-SOT TT-18 & 1  & 4.7 & 8.2 && 5.2 & 8.3    \\ 
t-SOT TT-18 & 2  & 4.4 & 7.6 && 5.0 & 7.6    \\
t-SOT TT-18 & 4  & 4.3 & 7.1 && 4.9 & 7.2    \\
t-SOT TT-18 & 8  & 4.3 & 6.9 && 4.9 & 7.0    \\
t-SOT TT-18 & 16 & 4.3 & 6.9 && 4.9 & 6.9    \\ \midrule
t-SOT TT-36 & 1  & 4.1 & 7.0 && 4.6 & 7.1    \\ 
t-SOT TT-36 & 2  & 4.0 & 6.5 && 4.5 & 6.8    \\
t-SOT TT-36 & 4  & 3.9 & 6.3 && 4.3 & 6.4    \\
t-SOT TT-36 & 8  & 3.9 & 6.1 && 4.3 & 6.2    \\ 
t-SOT TT-36 & 16 & 3.9 & 6.0 && 4.3 & 6.2    \\ \bottomrule
  \end{tabular}
  
  }
\end{table}

\section{LibriCSS experiment with high algorithmic latency}

Table \ref{tab:libricss_full} shows the evaluation result for LibriCSS based on t-SOT TT 
with different model configurations including one with high algorithmic latency of 2560 msec. 
As with the evaluation on LibriSpeechMix (Section \ref{sec:main_results}), 
we observed a significant WER reduction by increasing algorithmic latency,
and t-SOT TT-36 with 2560 msec of latency achieved SOTA results for all test conditions.

\begin{table}[h]
\ra{1.0}
  \caption{WER (\%) comparison on the monaural LibriCSS test set in the continuous input evaluation setting.
  A macro average of WERs is shown in the ``Avg.'' column.}
  \label{tab:libricss_full}
 \vspace{-3mm}
  \centering
{ \footnotesize 
  \tabcolsep = 1.6mm
\begin{tabular}{@{}lrrrrrrrr@{}}
    \toprule
Model & \multirow{2}{*}{\shortstack[c]{Latency\\(msec)}} &\multicolumn{7}{c}{WER (\%) for different overlap ratio} \\ \cmidrule{3-9}
        &  & 0L & 0S & 10 & 20 & 30 & 40 & Avg.   \\ \midrule
t-SOT TT-18 & 160  & 7.5 & 7.5 & 8.5 & 10.5  & 12.6  & 14.0 & 10.1 \\
t-SOT TT-36 & 160  & 6.7 & 6.1 & 7.5 & 9.3  & 11.6 & 12.9 & 9.0 \\
t-SOT TT-36 & 2560  & {\bf 5.4} & {\bf 5.3} & {\bf 6.5} & {\bf 7.3}  & {\bf 9.5} & {\bf 11.3} & {\bf 7.6} \\ \bottomrule
  \end{tabular}
  }
\end{table}

\end{document}